# Strong room-temperature blue-violet photoluminescence of multiferroic BaMnF$_4$

Shuang Zhou[1], Yakui Weng[1], Zhangting Wu[1], Jinlong Wang[2], Lingzhi Wu[3], Zhenhua Ni[1], Qingyu Xu[1,4,*], Shuai Dong[1,†]

[1] *Department of Physics, Southeast University, Nanjing 211189, & Key Laboratory of MEMS of the Ministry of Education, Southeast University, Nanjing 210096, & Collaborative Innovation Center of Suzhou Nano Science and Technology, Soochow University, Suzhou 215123, China*

[2] *School of Optoelectronic Engineering, Nanjing University of Posts and Telecommunications, Nanjing 210023, China*

[3] *School of Geography and Biological Information, Nanjing University of Posts and Telecommunications, Nanjing 210023, China*

[4] *National Laboratory of Solid State Microstructures, Nanjing University, Nanjing 210093, China*

**Abstract:**

BaMnF$_4$ microsheets have been prepared by hydrothermal method. Strong room-temperature blue-violet photoluminescence has been observed (absolute luminescence quantum yield 67%), with two peaks located at 385 nm and 410 nm, respectively. More interestingly, photon self-absorption phenomenon has been observed, leading to unusual abrupt drop of luminescence intensity at wavelength of 400 nm. To understand the underlying mechanism of such emitting, the electronic structure of BaMnF$_4$ has been studied by first principles calculations. The observed two peaks are attributed to electrons' transitions between the upper-Hubbard bands of Mn's $t_{2g}$ orbitals and the lower-Hubbard bands of Mn's $e_g$ orbitals. Those Mott gap mediated *d-d* orbital transitions may provide additional degrees of freedom to tune the photon generation and absorption in ferroelectrics.

**Keywords:** photoluminescence, fluoride, self-absorption, multiferroics

Corresponding authors: *E-mail:xuqingyu@seu.edu.cn; †E-mail:sdong@seu.edu.cn.

## 1. Introduction

Materials that can efficiently emit blue and violet light have attracted great research attentions in past decades due to their wide applications. For example, blue light-emitting materials are used in light-emitting diodes (LEDs) for full-color displays and in semiconductor laser diodes (LDs) for optical communication systems. Violet-light-emitting materials are used in digital versatile disks (DVDs) for higher storage capacity and in LDs for undersea optical communications. As we all know, the commonest inorganic fluorescent phosphors are alkaline earth metal sulfides (ZnS, CaS) [1-3] and aluminates ($SrAl_2O_4$, $CaAl_2O_4$ $BaAl_2O_4$) [4-6] as a matrix, with rare earth as activating agent or activator. Comparing with these two categories, solid inorganic fluorides have high optical transparency, lower phonon energy, high ionicity, electron-acceptor behavior and anionic conductivity [7-9], and also have wide range of promising optical applications in optics, biological labels and lenses [10, 11]. Most of inorganic fluorides phosphors need rare earth elements as activator. In addition, its complex preparation technology and toxic property hindered the application to a degree. By contrast, transition metal fluorides synthesized using a simple and low-cost fabrication process seem to be more economical and environmental. In this work, hydrothermal method was performed and the room-temperature photoluminescence (PL) of a selected inorganic multiferroic fluoride: $BaMnF_4$ has been studied both in experiments and first principles calculations.

$BaMnF_4$ belongs to the family of $BaMF_4$-type fluorides ($M$ = Mn, Fe, Co, Ni, Mg, Zn) which share the same orthorhombic structure. $BaMnF_4$ can be described by the non-centrosymmetric space group $A2_1am$, in which $Ba^{2+}$ ions are layered with sheets of distorted corners-sharing $[MnF_6]^{2-}$ octahedra, as shown in the inset of Fig. 1 [12]. It has attracted considerable research interests due to its multiferroic properties, which possesses a large spontaneous polarization along $a$ axis up to 11.5 μC/cm$^2$ [13] and antiferromagnetism with magnetic moment roughly along $b$ axis simultaneously [14, 15]. Such a multiferroic nature provides the possibility to tune $BMnF_4$'s physical properties via magnetic/electric stimulation. In addition, just like other inorganic fluorides, $BaMnF_4$ can also be applied for manufacturing of scintillators, high resolution color displays, white light-emission devices, security labels, monitoring equipment, cancer therapy drugs [16-18], etc.

Despite the intensively studied multiferroic properties, the investigation on optical properties of BaMnF$_4$ is rather rare [19], especially at room temperature. One possible reason is that the fluorides are much more difficult to be synthesized compared with widely studied oxides. In addition, BaMnF$_4$ has a structural phase transition occurring at 247 K and magnetic transition at 26 K [20, 21], which may influence the optical behaviors. Even though, the luminescence spectra of BaMnF$_4$ at wavelength of red and near infrared range were studied by Goldberg *et al.* more than thirty years ago. The emission bands at 600 nm, 640 nm, and 720 nm were observed at low temperature (10 K~110 K), but they did not mentioned its property at room temperature [21].

In this work, BaMnF$_4$ microsheets have been grown using hydrothermal method, and room-temperature PL in blue-violet region have been investigated when excited with ultra-violet light. Two strong emission bands have been clearly observed, their strongest positions locate at wavelength of 385 nm and the other at 410 nm, respectively. Photoluminescence excitation (PLE) spectra show a stable luminescent phenomenon for the two peaks. More interestingly, an unusual drop of luminescence intensity at 400 nm has been observed, which is conformed to be attributed to photon self-absorption effect. Furthermore, the underlying mechanism of PL has been explained as Mott-gaped *d-d* transition in according to first-principles electronic structure calculations.

**2. Experimental and methods**

Pale pink powder of BaMnF$_4$ microsheets was synthesized by hydrothermal method [22]. Stoichiometric BaF$_2$ and Mn(CH$_3$COO)$_2$·4H$_2$O were mixed, dissolved in trifluoroacetic acid solution (volume ratio of CF$_3$COOH and H$_2$O is 1:2) and the diluted solution was contained by polytetrafluoroethylene autoclave, heated to 220 $^o$C, held for 20 hours, and then cooled slowly to room temperature. After discarded the upper remaining liquid, the products was washed with ethanol, and dried in vacuum condition. All the reagents were used as starting materials without further purification.

The equipment were used and characterized the structure and morphology of the prepared sample are as following: X-ray diffraction (XRD, Rigaku Smartlab3) with Cu *Kα* radiation, transmission electron microscope (TEM, Tecnai F20) and scanning electron microscope (SEM, FEI Inspection F50). PL and PLE were measured by using spectrofluorometer (fluorolog3-TCSPC, Horiba Jobin Yvon). Absorption measurement was carried with spectrophotometer (UV-3600), and fluorescence efficiency (characterized as absolute quantum yield) was recorded on an Edinburgh FLS920P spectrometer with a integrating sphere.

The first-principles density-function theory (DFT) calculations were performed using the spin-polarized local density approximation (LDA) method with Hubbard $U$ correction, based on the projector-augmented wave (PAW) potentials, as implemented in the Vienna *ab initio* Simulation Package (VASP) [23, 24]. Various values of the effective Hubbard coefficient ($U_{eff}=U-J$) on Mn's $3d$ states has been tested from 0 eV to 4 eV [25-27]. The cutoff energy of plane-wave is 550 eV and the $k$-point mesh is $\Gamma$-centered $5\times7\times5$. The experimental lattice constants and internal atomic positions are adopted in the following calculation as the initial values [20], which are fully optimized till the Hellman-Feynman forces are converged to less than 0.01 eV/Å. The experimental antiferromagnetism is adopted [28].

## 3. Results and discussion

The structural and sample quality of our BaMnF$_4$ powder are checked by XRD, as shown in Fig. 1, which confirms the orthorhombic structure with space group of *A2$_1$am*. No impurity phase can be detected from the XRD pattern. Schematic crystal structure of BaMnF$_4$ is sketched in the inset plot where the sheet structure consists of distorted corners-sharing [MnF$_6$]$^{2-}$ octahedra. The morphology of our microcrystals was studied using SEM images with different magnifications as shown in Fig. 2(a-b), the banded sheets have regular shape and lateral size is of several micrometers. This micro-sheet morphology suggests the anisotropic growth under hydrothermal condition, which should be related to the layered-like crystal structure. The inset of Fig. 2(b) shows selected area electron diffraction (SAED) pattern taken in TEM, which is a good evidence for the crystals to be micro-sized single crystal with good quality.

BaMnF$_4$ powder was pressed into thin circular tablet for luminescence measurements.

During the measurements for PL and PLE, small gratings of 2 nm were used for both incident and emergent light detectors. Figure 3(a) shows the PL spectra with various excitation wavelengths ($\lambda_{exc}$=260-360 nm). All these spectra show similar shapes, containing two main emission bands. These two strongest emission positions locate at 385 nm and 410 nm, respectively. The peak positions of luminescence spectra are robust, almost unchanged when the excitation wavelength is tuned. With various excitation light, the PL spectrum excited by 280 nm ultra-violet light has the strongest intensity, i.e. the optimal exciting wavelength. Both two emission peaks have large linewidth which may be due to broad distribution of the particle size or strong electron-phonon coupling of multiferroics [29]. Furthermore, the fluorescence emission spectrum at 78 K (inset of Fig. 3(a)) shows the same behavior with that at room temperature, i.e. two peaks and a valley at the identical positions, which suggests such a luminescence behavior of $BaMnF_4$ to be temperature-independent, at least between 78 K and room temperature.

Figure 3(b) shows the PLE spectra of $BaMnF_4$ excited by the light of wavelengths in the range of 235 and 340 nm. The emission intensities at wavelengths of both 385 nm and 410 nm were monitored. In whole, the PLE intensity at 385 nm is higher than that at 410 nm, in consistent with aforementioned PL spectra. The inset of Fig. 3(b) shows the luminescent photo image of $BaMnF_4$ powder excited by the ultraviolet light. The bright blue color is due to the selectivity of naked eyes. The fluorescence efficient of $BaMnF_4$ was measured by using integrating sphere, and its absolute luminescence quantum yield (300nm - 520nm) reaches ~67%, and this is quite a high value for inorganic fluorides phosphors.

In contrast to the general overlapping shape of two neighboring emission bands [30], a sudden drop of PL intensity at wavelength of 400 nm can be clearly evidenced in Fig. 3(a), which may due to photon self-absorption mechanism [31], and this will be discussed later. This absorption is robust and unvaried with the excitation wavelengths in the range of 260 nm and 360 nm.

The proposed interpretation of luminescence can be further checked by the first-principles calculations. By varying the Hubbard coefficient, it is found that $U_{eff}$=1 eV gives the best description of $BaMnF_4$. The local magnetic moment within the Wigner-Seitz sphere is 4.49 $\mu_B$ per Mn atom, implying the high-spin state for $Mn^{2+}$, as expected. The density of state (DOS) and projected density of states (PDOS) show that electronic bands near the Fermi level are from Mn's $3d$ orbitals, as shown in Fig. 3(c). A nontrivial character of $BaMnF_4$'s electronic structure is that both the $t_{2g}$ and $e_g$ bands are very narrow, implying extremely localized $3d$ state. This localization is partially due to the half-filling fact of Mn's $3d$ orbital which is the most ideal condition for Mottness. Another reason is the weak hybridization between Mn's $3d$ orbitals and F's $2p$ orbitals whose energy is much lower.

The band gap of $BaMnF_4$ in our DFT calculation is about 3.0 eV (when $U_{eff}$=1 eV), separating the empty upper-Hubbard $t_{2g}$ bands and occupied lower-Hubbard $e_g$ bands. This Mott gap coincides with the emission photon energy, suggesting a $d$-$d$ transition induced PL. It is well known that the intrashell photon excitation/emission is usually quite weak due to the selection rule. In this sense, however, the quantum yield of $BaMnF_4$ is as high as 67% (300 nm - 520 nm), suggesting that the strong room-temperature PL of $BaMnF_4$ observed here is quite prominent considering the fact that here no any rare earth element is involved. Noting that the spontaneous ferroelectric polarization of $BaMnF_4$ distorts the lattice and changes the symmetry of electronic wavefunctions, which may be responsible for the violation of selection rule as proposed for ideal isolated atoms.

To reveal more information of the sudden drop of intensity at 400 nm, the absorption spectrum of $BaMnF_4$ powder was measured ranging from 300 nm to 550 nm, as shown in Fig. 4(a). The conspicuous peaks (A-E) of absorption are summarized in Tab. 1. Our result of absorption agrees with previous unpolarized absorption spectrum of $BaMnF_4$ single crystal who also shows a sharp peak at 400 nm at 295 K [31]. Compare the peaks of our experimental and the reported ones in Ref. [32], it is clear that all the peaks nearly have the same photon energy. This is also a powerful evidence for the sample quality of $BaMnF_4$. Furthermore, Fig.4 (b) shows the contrast figure between PL spectrum excited with 280 nm (red) and inverted absorption (green). It is obvious that the strong absorption peak of 400 nm fits well with the deep valley of PL, both for the same shape and exact position. In addition, PLE in Fig. 3 (b) presents the two peaks' similar phenomenon. Till now, the supposed self-absorption has been conformed rationally. Self-absorption is a common case for solid luminescent materials, and has been researched for many years [31, 33-37]. In some solid state, physical phenomena such as multiple scattering and self-absorption of the emitted light may occur, leading to the distortion or split of the luminescence features, thus compromising the data interpretation [31, 38]. In our system, the two detached peaks in PL spectra should originally belong to a broad peak with large line width. However, because of the self-absorption effect in 400 nm, the broad peak was divided into two parts.

According to the DFT band structure (Fig. 3(d)), both the $e_g$ bands and $t_{2g}$ bands are further split due to the Jahn-Teller distortion of Mn-F$_6$ octahedra. Especially for the $e_g$ ones, the split is about 0.1 eV. By considering such splitting, the two peaks of emission and the self-absorption can be well mapped to the transitions among these sub-bands, as indicated in Fig. 3(d). First, the electrons are excited to the unoccupied upper-Hubbard $t_{2g}$ orbitals by the exciting ultraviolet light, then the transition of excited electrons from unoccupied $t_{2g}$ orbitals to the split $e_g$ orbitals leads to the PL emissions at 385 nm and 410 nm. The emission photons can be absorbed and the electrons at the upper $e_g$ orbitals are excited to the upper unoccupied $t_{2g}$ orbitals, leading to the deep valley at 400 nm of PL spectra.

## 4. Conclusions and perspective

BaMnF$_4$ microsheets were synthesized using the hydrothermal method, and the PL and PLE spectra at room temperature have been studied. Strong emissions at 385 nm and 410 nm can be excited by the ultraviolet illuminations with wavelength ranging from 260 nm to 340 nm, and 280 nm is the most efficient excitation. By comparing absorption spectra and PL spectra, it is conformed that the sudden drop of PL spectra at 400 nm was induced by self-absorption effect. According to first-principles calculations, the emissions are mainly due to the *d-d* transitions between the split t$_{2g}$ and e$_g$ orbitals by both the Hubbard repulsion as well as the Jahn-Teller distortion of [MnF$_6$]$^{2-}$ octahedra in BaMnF$_4$. Since both the Hubbard bands and Jahn-Teller distortion can be tuned by many methods, e.g. doping or strain, the strong room-temperature PL of BaMnF$_4$ may be tunable for better applications, which will be studied in the future.

**Acknowledgments**

This work is supported by the National Natural Science Foundation of China (51172044, 51471085, 51322206), the Natural Science Foundation of Jiangsu Province of China (BK20151400).

**References:**

1. J. K. Cooper, S. Gul, S. A. Lindley, J. Yano and J. Z. Zhang, *ACS Appl. Mater. Inter.*, 2015, **7**, 10055.
2. X. L. Wang, J. Y. Shi, Z. C. Feng, M. R. Li and C. Li, *Phys. Chem. Chem. Phys.*, 2011, **13**, 4715.
3. M. Georgin, L. Carlini, D. Cooper, S. E. Bradforth and J. L. Nadeau, *Phys. Chem. Chem. Phys.*, 2013, **15**, 10418.
4. A. A. Yaremchenko, V. V. Kharton, A. A. Valente, S. A. Veniaminov, V. D. Belyaev, V. A. Sobyanin and F. M. B. Marques, *Phys. Chem. Chem. Phys.*, 2007, **9**, 2744.
5. F. Clabau, X. Rocquefelte, S. Jobic, P. Deniard, M. H. Whangbo, A. Garcia and T. Le Mercier, *Chem. Mater.*, 2005, **17**, 3904.
6. S. H. Jua, U. S. Oha, J. C. Choia, H. L. Parka, T. W. Kimb and C. D. Kimc, *Mater. Res. Bull.*, 2000, **35**, 1831.
7. C. Feldmann, M. Roming and K. Trampert, *Small*, 2006, **2**, 1248.
8. Z. W. Quan, D. M. Yang, P. P. Yang, X. M. Zhang, H. Z. Lian, X. M. Liu and J. Lin, *Inorg. Chem.*, 2008, *47*, 9509.
9. P. Gao, Y. Xie and Z. Li, *Eur. J. Inorg. Chem.*, 2006, **16**, 3261.
10. Y. B. Mao, F. Zhang and S. S. Wong, *Adv. Mater.*, 2006, **18**, 1895.
11. W. S. Wang, L. Zhen, C. Y. Xu, J. Z. Chen and W. Z. Shao, *Appl. Mater. Interfaces*, 2009, **1**, 780.
12. C. Ederer and N. Spaldin, *Phys. Rev. B*, 2006, **74**, 024102.
13. L. F. David and J. F. Scott, *J. Phys. C: Solid State Phys.*, 1977, **10**, L329.
14. P. Sciau, M. Clin, J. P. Rivera and H. Schmid, *Ferroelectrics*, 1989, **105**, 201.
15. D. E. Cox, S. M. Shapiro, M. Eibschiitz, H. J. Guggenheim and R. A. Cowley, *Phys. Rev. B*, 1979, **19**, 5754.

16. M. C. Marco de Lucas, M. Moreno, F. Rodriguez and P. G. Baranov, *J. Phys.: Condens. Matter.*, 1996, **8**, 2457.
17. S. F. Lim, W. S. Ryu and R. H. Austin, *Opti. Express*, 2010, **18**, 2309.
18. X. Yang, S. Xiao, J. W. Ding and X. H. Yan, *J. Mater. Sci.*, 2007, **42**, 7042.
19. C. Li and J. Lin, *J. Mater. Chem.*, 2010, **20**, 6831.
20. V. Franco-Puntes, K. M. Krishnan and A. P. Alivisatos, *Science*, 2001, **291**, 2115.
21. V. Goldberg, D. Pacheco, R. Moncorge and B. Di Bartolo, *J. Lumin.*, 1979, **18**, 143.
22. S. W. Kim, H. Y. Chang and P. S. Halasyamani, *J. Am. Chem. Soc.*, 2010, **132**, 17684.
23. G. Kresse and J. Hafner, *J. Non-cryst. Solids*, 1995, **193**, 222.
24. G. Kresse and J. Furthmuller, *Phys. Rev. B*, 1996, **54**, 11169.
25. M. Cococcioni and S. D. Gironcoli, *Phys. Rev. B*, 2005, **71**, 035105.
26. T. Hashimoto, S. Ishibashi and K. Terakura, *Phys. Rev. B*, 2010, **82**, 079903.
27. S. Dong, W. Li, X. Huang and E. Dagotto, *J. Appl. Phys.*, 2014, **115**, 17D723.
28. D. E. Cox, M. Eibschutz, H. J. Guggenheim and L. Holmes, *J. Appl. Phys.*, 1970, **41**, 943.
29. A. Endoh, Y. Nakata, Y. Sugiyama, M. Takatsu and N. Yokoyama, *Jpn. J. Appl. Phys.*, 1998, **38**, 1085.
30. L. Liu, D. M. Zhang, Y. S. Zhang, Z. G. Bao and L. F. Chen, *J. Mater. Sci.*, 2013, **48**, 876.
31. C. Clementi, C. Miliani, G. Verri, S. Sotiropoulou, A. Romani, B. G. Brunetti and A. Sgamellotti, *Appl. Spectrosc.*, 2009, **63**, 1323.
32. T. Tsuboi and W. Kleemann, *Phys. Rev. B*, 1983, **27**, 3762.
33. C. Clementi, F. Rosi, A. Romani, R. Vivani, B. G. Brunetti and C. Miliani, *Appl. Spectrosc.*, 2012, **66**, 1233.
34. K. Suzuki, A. Kobayashi, S. Kaneko, K. Takehira, T. Yoshihara, H. Ishida, Y. Shiina, S. Oishi and S. Tobita, *Phys. Chem. Chem. Phys.*, 2009, **11**, 9850.
35. T. S. Ahn, R. O. Al-Kaysi, A. M. Muller, K. M. Wentz and C. J. Bardeen, *Rev. Sci. Instrum.*, 2007, **78**, 086105.
36. N. Kaihovirta, G. Longo, L. Gil-Escrig, H. J. Bolink and L. Edman, *Appl. Phys. Lett.*, 2015, ***106***, 103502.

37. G. Verri, C. Clementi, D. Comelli, S. Cather and F. Pique, *Appl. Spectrosc.*, 2008, **62**, 1295.
38. T. Suhasini, B. C. Jamalaiah, T. Chengaiah, S. J. Kumar and R. L. Moorthy, *Physica B*, 2012, **407**, 523.

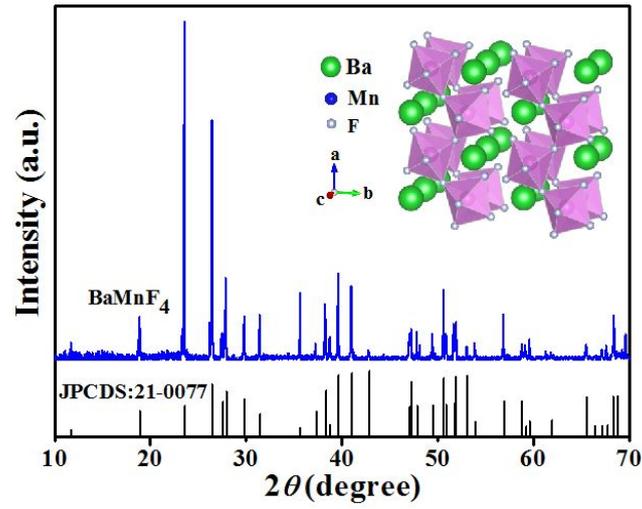

**Figure 1.** XRD pattern of BaMnF$_4$ microsheets synthesized by the hydrothermal method (main plot). The schematic crystal structure of BaMnF$_4$ (the inset).

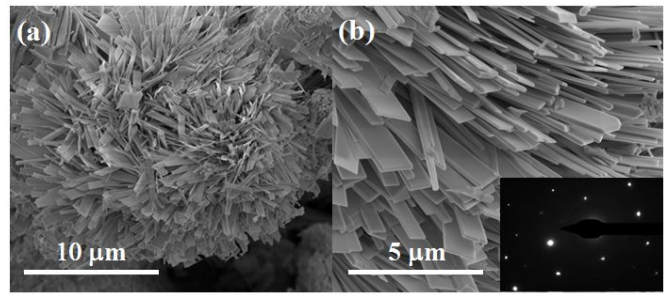

**Figure 2.** (a-b) SEM images with different magnifications of BaMnF$_4$ microsheets. Inset of (b): SAED pattern of single crystal taken in TEM.

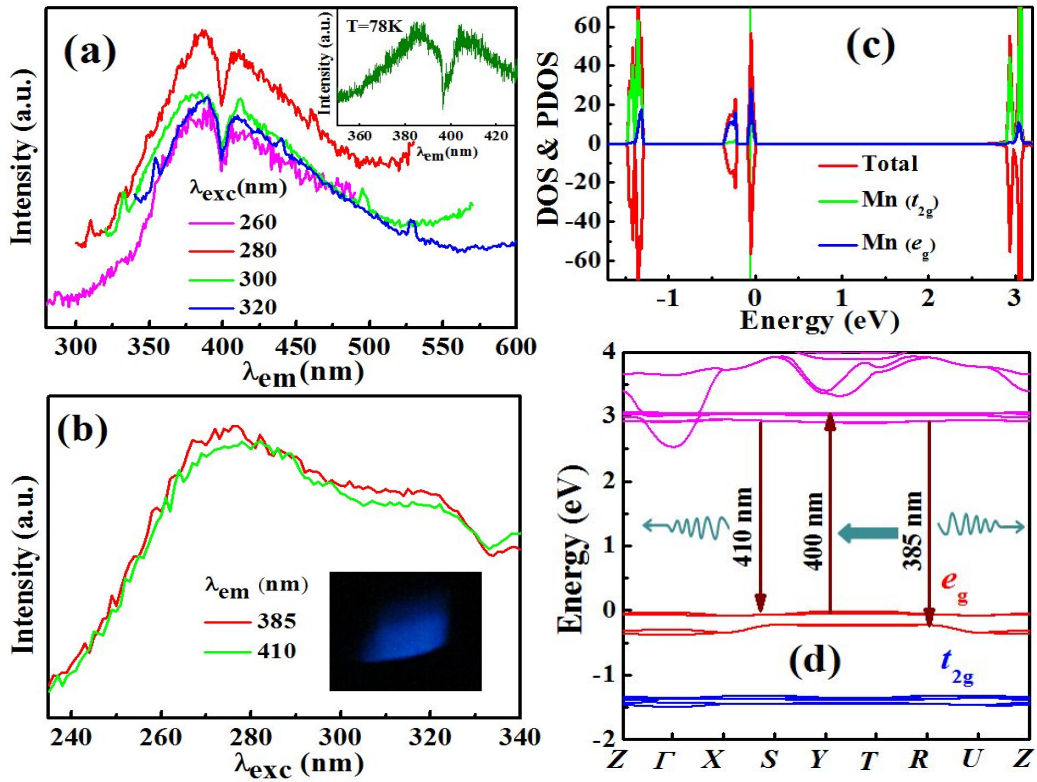

**Figure 3.** (a) PL spectra of BaMnF$_4$ under various excitations; Inset shows the PL spectra excited by 325 nm laser at 78 K; (b) PLE spectra with emission wavelength of 385 nm and 410 nm; Inset shows the digital photo of BaMnF$_4$ powder under ultraviolet illumination; (c) DOS of BaMnF$_4$. The orbital-resolved PDOS is also calculated and the five 3$d$ orbitals are grouped into two categories (the triplet $t_{2g}$: $d_{xy}$, $d_{yz}$, $d_{xz}$; and the doublet $e_g$: $d_{x2-y2}$, $d_{3z2-r2}$) . (d) The band structures. The arrows denote the mechanism of PL emission and self-absorption.

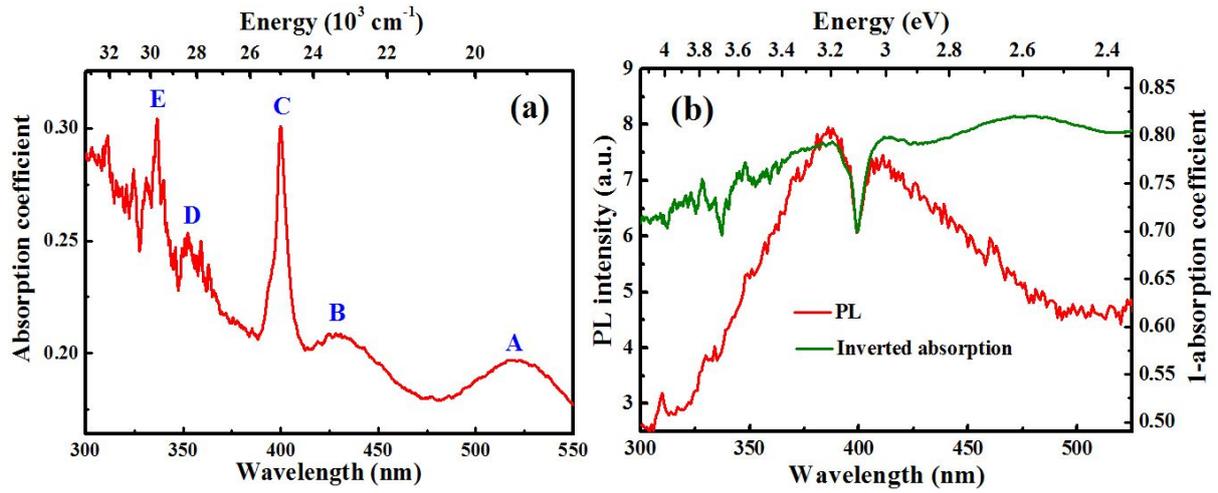

**Figure 4.** (a) Absorption spectra of BaMnF$_4$. (b) Comparison between PL spectrum excited with 280 nm (red) and inverted absorption (green).

**Table 1.** Information of the main peaks A-E corresponding to the absorption spectrum in Fig. 4(a). The data presented in the second column are taken from Ref. 32.

| Peak | Photon energy (10$^3$ cm$^{-1}$) [32] | Photon energy (10$^3$ cm$^{-1}$) | Wavelength (nm) | Energy (eV) |
|---|---|---|---|---|
| A | 19.15 | 19.20 | 520.8 | 2.38 |
| B | 23.05 | 23.18 | 431.4 | 2.87 |
| C | 25.22 | 25.00 | 400.0 | 3.10 |
| D | 28.00 | 28.33 | 353.0 | 3.51 |
| E | 30.05 | 29.68 | 336.9 | 3.68 |